\documentclass[twocolumn]{aastex61}

\hypersetup{citecolor=blue,urlcolor=blue}

\received{June 22, 2017}
\accepted{July 18, 2017}

\shorttitle{On the disappearance of a torus in NGC 1097}
\shortauthors{T. Izumi et al.}

\begin{document}
\title{On the disappearance of a cold molecular torus\\ around the low-luminosity active galactic nucleus of NGC 1097}

\correspondingauthor{Takuma Izumi}
\email{takuma.izumi@nao.ac.jp}

\author[0000-0002-0786-7307]{T. Izumi}
\affil{National Astronomical Observatory of Japan, 2-21-1 Osawa, Mitaka, Tokyo 181-8588, Japan}

\author{K. Kohno}
\affil{Institute of Astronomy, Graduate School of Science, The University of Tokyo, 2-21-1 Osawa, Mitaka, Tokyo 181-0015, Japan}
\affil{Research Center for the Early Universe, Graduate School of Science, The University of Tokyo, 7-3-1 Hongo, Bunkyo, Tokyo 113-0033, Japan}

\author{K. Fathi}
\affil{Valhallav. 67, 114 28 Stockholm, Sweden}

\author{E. Hatziminaoglou}
\affil{European Southern Observatory, Karl-Schwarzschild-Str. 2, 85748 Garching bei M\"{u}nchen, Germany}

\author{R. I. Davies}
\affil{Max-Planck-Institute f\"{u}r Extraterrestrische Physik, Postfach 1312, D-85741 Garching, Germany}

\author{S. Mart\'{i}n}
\affil{Joint ALMA Observatory, Alonso de C\'{o}rdova 3107, Vitacura, Santiago 763-0355, Chile}
\affil{European Southern Observatory, Alonso de C\'{o}rdova 3107, Vitacura, Santiago 763-0355, Chile}

\author{S. Matsushita}
\affil{Academia Sinica, Institute of Astronomy \& Astrophysics, P.O. Box 23-141, Taipei 10617, Taiwan}

\author{E. Schinnerer}
\affil{Max Planck Institute for Astronomy, K\"{o}nigstuhl 17, Heidelberg 69117, Germany}

\author{D. Espada}
\affil{National Astronomical Observatory of Japan, 2-21-1 Osawa, Mitaka, Tokyo 181-8588, Japan}
\affil{SOKENDAI (The Graduate University for Advanced Studies), 2-21-1 Osawa, Mitaka, Tokyo 181-8588, Japan}

\author{S. Aalto} 
\affil{Department of Space, Earth and Environment, Chalmers University of Technology, Onsala Observatory, SE-439 92 Onsala, Sweden}

\author{K. Onishi}
\affil{Research Center for Space and Cosmic Evolution, Ehime University, 2-5 Bunkyocho, Matsuyama, Ehime 790-0826, Japan}

\author{J. L. Turner}
\affil{Department of Physics and Astronomy, UCLA, 430 Portola Plaza, Los Angeles, CA 90095-1547, USA}

\author{M. Imanishi}
\affil{National Astronomical Observatory of Japan, 2-21-1 Osawa, Mitaka, Tokyo 181-8588, Japan}
\affil{SOKENDAI (The Graduate University for Advanced Studies), 2-21-1 Osawa, Mitaka, Tokyo 181-8588, Japan}
\affil{Subaru Telescope, NAOJ, 650 North A’ohoku Place, Hilo, HI 96720, USA}

\author{K. Nakanishi}
\affil{National Astronomical Observatory of Japan, 2-21-1 Osawa, Mitaka, Tokyo 181-8588, Japan}
\affil{SOKENDAI (The Graduate University for Advanced Studies), 2-21-1 Osawa, Mitaka, Tokyo 181-8588, Japan}

\author{D. S. Meier}
\affil{Department of Physics, New Mexico Institute of Mining and Technology, 801 Leroy Place, Soccoro, NM 87801, USA}

\author{K. Wada}
\affil{Kagoshima University, Kagoshima 890-0065, Japan}

\author{N. Kawakatu}
\affil{Faculty of Natural Sciences, National Institute of Technology, Kure College, 2-2-11 Agaminami, Kure, Hiroshima 737-8506, Japan}

\author{T. Nakajima}
\affil{Institute for Space-Earth Environmental Research, Nagoya University, Furo-cho, Chikusa-ku, Nagoya, Aichi 464-8601, Japan}



\begin{abstract}
We used the Atacama Large Millimeter/submillimeter Array (ALMA) 
to map the CO(3--2) and the underlying continuum emissions 
around the type 1 low-luminosity active galactic nucleus 
(LLAGN; bolometric luminosity $\lesssim 10^{42}$ erg~s$^{-1}$) 
of NGC 1097 at $\sim 10$ pc resolution. 
These observations revealed a detailed cold gas distribution 
within a $\sim 100$ pc of this LLAGN. 
In contrast to the luminous Seyfert galaxy NGC 1068, 
where a $\sim 7$ pc cold molecular torus was recently revealed, 
a distinctively dense and compact torus is missing in 
our CO(3--2) integrated intensity map of NGC 1097. 
Based on the CO(3--2) flux, the gas mass of the torus of NGC 1097 
would be a factor of $\gtrsim 2-3$ less than that found for NGC 1068 
by using the same CO-to-H$_2$ conversion factor, 
which implies less active nuclear star formation and/or inflows in NGC 1097. 
Our dynamical modeling of the CO(3--2) velocity field implies that 
the cold molecular gas is concentrated in a thin layer as compared to 
the hot gas traced by the 2.12 $\mu$m H$_2$ emission in and around the torus. 
Furthermore, we suggest that NGC 1097 hosts 
a geometrically thinner torus than NGC 1068. 
Although the physical origin of the torus thickness remains unclear, 
our observations support a theoretical prediction that 
geometrically thick tori with high opacity will become deficient 
as AGNs evolve from luminous Seyferts to LLAGNs. 
\end{abstract}


\section{Introduction}\label{sec1}
The unified scheme of active galactic nuclei (AGNs) 
postulates that the existence (type-1) or absence (type-2) of a broad line region 
depends on the viewing angle of an optically and geometrically thick 
dusty/molecular torus \citep{1993ARA&A..31..473A}. 
Spatially resolved thermal dust emissions at near- to mid-infrared (NIR to MIR) wavelengths 
in AGNs support the existence of compact ($< 10$ pc) obscuring structures 
\citep[e.g.,][]{2004Natur.429...47J,2013A&A...558A.149B}. 

In addition to the photometric evidence, \citet{2009ApJ...696..448H} unveiled 
10 pc scale molecular disks with high velocity dispersions 
through dynamical modeling of the 2.12 $\micron$ H$_2$ 
(hereafter NIR H$_2$) emission line in local AGNs, 
which would correspond to the outer envelope of tori. 
These high velocity dispersion regions were also observed 
in NGC 1068 \citep{2016ApJ...822L..10I,2016ApJ...823L..12G,2016ApJ...829L...7G}, 
NGC 1377 \citep{2017arXiv170205458A}, and Centaurus A \citep[Cen A;][]{2017arXiv170605762E} 
with the Atacama Large Millimeter/submillimeter Array (ALMA), 
in cold gas and dust emission. 
However, the physical origin of the high velocity dispersion 
or the vertical height of a torus is still being debated. 

Some models suggest that a torus is an evolving structure 
energetically driven by, for example, massive inflows or a nuclear starburst 
\citep[e.g.,][]{2008A&A...491..441V,2009ApJ...702...63W}, 
both of which can ignite the AGN activity. 
If those triggers become weak, 
the geometric thickness of the torus cannot be sustained. 
Disk winds \citep{2006ApJ...648L.101E,2016PASJ...68...16N} 
or outflows driven by the AGN radiation \citep{2012ApJ...758...66W} 
themselves will also generate a toroidal structure. 
In either case, the geometry and opacity of the torus would depend on the AGN luminosity. 
This prompted us to investigate low-luminosity AGNs 
\citep[LLAGN, bolometric luminosity $L_{\rm bol} \lesssim 10^{42}$ erg s$^{-1}$;][]{2008ARA&A..46..475H}, 
where thick tori may be {\it deficient} due to insufficient energy input. 
Indeed, theories predict such deficiency at this LLAGN regime \citep[e.g.,][]{2006ApJ...648L.101E}. 
Recent clumpy torus modelings of infrared spectral energy distributions \citep{2015A&A...578A..74G,2017ApJ...841...37G}, 
as well as kinematic modelings of the NIR H$_2$ emission line \citep{2013ApJ...763L...1M}, 
of AGNs with various $L_{\rm bol}$ supported the trend that 
LLAGNs have geometrically and optically thinner tori than luminous Seyferts. 

Here, we present our 10 pc diameter resolution ALMA cycle 3 observations 
of CO(3--2) ($\nu_{\rm rest}$ = 345.796 GHz) 
and the underlying continuum emissions 
toward the central $\sim 100$ pc circumnuclear disk (CND) region of the nearby 
type-1 \citep{1993ApJ...410L..11S} LLAGN of 
NGC 1097 \citep[$D$ = 14.5 Mpc, 1$\arcsec$ = 70 pc;][]{1988ngc..book.....T}. 
Considering its low-luminosity \citep[$L_{\rm bol}$ = 8.6 $\times$ 10$^{41}$ erg s$^{-1}$;][]{2006ApJ...643..652N}, 
a comparison of observed torus properties of this LLAGN 
with those of luminous Seyferts such as NGC 1068 
\citep[$L_{\rm bol}$ = 2 $\times$ 10$^{45}$ erg s$^{-1}$;][]{2016MNRAS.456L..94M} 
will shed light on the torus evolution. 
Given that the NIR H$_2$ emission has complex excitation mechanisms 
and may reflect a small fraction of the molecular mass, 
cold gas observations could be beneficial for this kind of comparison. 
Note that the dynamical mass ($M_{\rm dyn}$) at $r < 30$ pc is $\times 2.5$ higher 
in NGC 1068 than in NGC 1097 \citep{2009ApJ...696..448H}. 
Furthermore, based on high resolution ALMA cycle 1 data, Hatziminaoglou et al. (in prep.) 
found that the black hole mass ($M_{\rm BH}$) of NGC 1097 to be smaller 
than that \citep[$\sim 10^7~M_\Sun$;][]{1996ApJ...472L..21G} of NGC 1068. 
Therefore, if the torus thickness is solely determined by the disk gravity, 
we would expect a thinner torus in NGC 1068 than in NGC 1097, 
which contrasts with the evolutionary torus models.

\section{Observations}\label{sec2} 
Our ALMA cycle 3 observations consisted of four executions 
using the band 7 receiver (ID \#2015.1.00126.S) 
with 38--39 antennas in September 2016, 
with a total on-source time of 2.8 h. 
One of the four spectral windows (each with a width of 1.875 GHz) was used to fully cover the CO(3--2) emission line. 
The projected baselines ranged from 15.1 m to 3.1 km, 
resulting in a maximum recoverable scale of $\sim 7\arcsec$. 
The reduction and calibration were conducted with CASA version 4.7 
\citep{2007ASPC..376..127M} in the standard manner. 
To improve the image fidelity of the CO(3--2) cube, 
we combined cycle 3 and cycle 1 visibilities (ID \#2012.1.00187.S). 
The baseline of the cycle 1 data (three executions) 
ranged from 15.1 m to 1.6 km with 33--49 antennas. 
The total on-source time of the combined cube is 4.0 h. 
Continuum emission was subtracted 
individually from each execution data in the $uv$-plane. 

All images presented in this paper were 
reconstructed using the \verb|CLEAN| task 
with Briggs weighting (robust = 0.5). 
The resultant synthesized beams were 
0\arcsec.17 $\times$ 0\arcsec.11 (P.A. = 88$\arcdeg$.8) for the combined CO(3--2) cube, 
$\sim 0\arcsec.1$ for the cycle 3 continuum maps ($\nu_{\rm rest}$ = 350.3 GHz), 
and $\sim 0\arcsec.2-0\arcsec.25$ for the cycle 1 continuum maps, 
which correspond to $\sim 7-15$ pc. 
The 1$\sigma$ sensitivities were 0.84 mJy beam$^{-1}$ 
for the combined CO(3--2) cube (velocity resolution $dV$ = 10 km s$^{-1}$), 
45--72 $\mu$Jy beam$^{-1}$ for the cycle 3 continuum maps, 
and 70--110 $\mu$Jy beam$^{-1}$ for the cycle 1 continuum maps. 
The adopted absolute flux uncertainty was $\sim 10\%$ 
according to the ALMA Cycle 3 Proposer's Guide, 
which is included in our quantitative results. 
Although we lost $\sim 75\%$ of the CO(3--2) flux as compared to 
the 18$\arcsec$ aperture Atacama Pathfinder EXperiment (APEX) 
telescope \citep{2011MNRAS.414..529P}, 
this would have had little impact on our study 
as we focused on the central 1$\arcsec$. 
Our combined cubes were further 
analyzed with MIRIAD \citep{1995ASPC...77..433S}.

\section{Result}\label{sec3}
\subsection{Continuum emission}\label{sec3.1}

\begin{figure}[h]
\epsscale{1.1}
\plotone{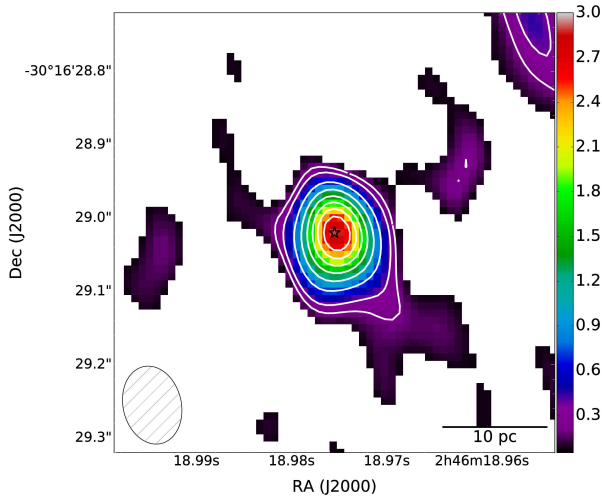}
\caption{
Band 7 ($\nu_{\rm rest}$ = 350.3 GHz) continuum 
emission at the central region of NGC 1097, 
shown in both the color scale (mJy beam$^{-1}$ unit) 
and contours (3, 5, 10, 15, $...$, 40$\sigma$, 
where 1$\sigma$ = 72.2 $\mu$Jy beam$^{-1}$). 
This map is made with the data obtained on September 14, 2016, 
as the brightest example of the seven continuum 
measurements at different dates (see also Figure \ref{fig2}). 
The bottom-left ellipse indicates the synthesized beam 
(0\arcsec.11 $\times$ 0\arcsec.08, P.A. = 13$\arcdeg$.0). 
The star marks the AGN position. 
}
\label{fig1}
\end{figure}

Figure \ref{fig1} shows the spatial distribution 
of the ALMA band 7 continuum emission at the central $\sim 40$ pc of NGC 1097. 
This is made from the one cycle 3 execution data 
(14 September, 2016; 0$\arcsec$.11 $\times$ 0$\arcsec$.08 resolution). 
The 350 GHz continuum emission peaks at 
$\alpha_{\rm J2000.0}$ = 02$^{\rm h}$46$^{\rm m}$18$^{\rm s}$.9755, 
$\delta_{\rm J2000.0}$ = -30$\arcdeg$16$\arcmin$29$\arcsec$.02, 
which coincides precisely with the 14.9 GHz peak position identified 
with a 1$\arcsec$.15 $\times$ 0$\arcsec$.45 beam \citep{2010MNRAS.401.2599O}. 
Thus we define this as the AGN location. 

A Gaussian fit to the visibilities with the MIRIAD task \verb|uvfit| 
revealed that this emission was unresolved at our $\sim 7$ pc resolution. 
This is in contrast to the case of NGC 1068, 
in which a dusty torus with a $\sim 7-10$ pc 
extent (diameter) was identified at 266 GHz and 694 GHz continuum emission 
\citep{2016ApJ...822L..10I,2016ApJ...823L..12G,2016ApJ...829L...7G}. 
A dusty torus in NGC 1097 must have a very low surface brightness 
at 350 GHz or be compact compared with the beam size. 

\begin{figure}[h]
\epsscale{1.1}
\plotone{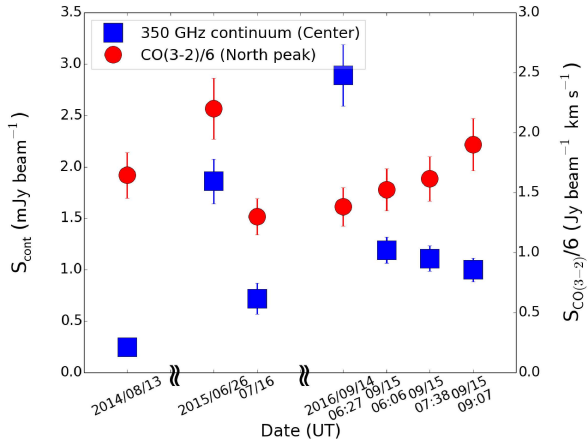}
\caption{
Time variation of the $\nu_{\rm rest}$ = 350.3 GHz continuum emission 
at the central $r \lesssim 9$ pc of NGC 1097. 
The horizontal axis indicates the starting date of each execution. 
The CO(3--2) integrated intensities 
(matched $uv$ range: 40 to 1480 k$\lambda$; matched resolution = 0$\arcsec$.3) 
measured at the north-knot (Figure \ref{fig3}c) are shown as the references. 
}
\label{fig2}
\end{figure}

We found significant time variability of the 350 GHz continuum flux level 
in the central $r \lesssim 9$ pc during our seven executions (Figure \ref{fig2}). 
This is genuine as 
(i) the variability amplitude (factor $\sim 2-5$ around the mean = 1.3 mJy) 
was well beyond the absolute flux uncertainty 
and (ii) the simultaneously observed CO(3--2) flux 
remained almost constant over the executions. 
Difference in apertures between cycles 1 and 3 does not influence the result 
as the continuum emission is unresolved. 
Therefore, the band 7 continuum is dominated by this time-varying 
component, which seems to be the non-thermal synchrotron emission 
of a radio jet according to the spectral energy distribution modeling by Hatziminaoglou et al. (in prep.). 

An upper limit on thermal dust emission then 
comes from our {\it lowest} continuum flux, i.e., 0.25 $\pm$ 0.08 mJy. 
Following the method in \S 3 of \citet{2016ApJ...823L..12G}, 
we estimated a mass of corresponding cold dust $M_{\rm dust}$ $<$ 530 $\pm$ 30 $M_\Sun$ 
within a single beam (0$\arcsec$.26 $\times$ 0$\arcsec$.13) of that observation, 
where we assumed a modified black body spectrum with an emissivity index $\beta = 2$ 
and a dust emissivity $\kappa_{\rm 352 GHz} = 0.0865$ m$^2$~kg$^{-1}$ 
\citep{2001A&A...379..823K}. 
A dust temperature of 150 K was assumed, which is a representative 
molecular gas kinetic temperature at the central $r < 120$ pc \citep{2013PASJ...65..100I}. 
If we decrease the dust temperature by 50\%, 
$M_{\rm dust}$ is accordingly increased by $\sim 50$\% 
as it is close to the Rayleigh-Jeans regime. 
The time-variable nature clearly indicates that one should avoid 
using a single continuum measurement at the band 7 (and lower frequency) to estimate $M_{\rm dust}$. 

\subsection{CO(3--2) emission}\label{sec3.2}
Figure \ref{fig3}a shows the global distribution 
of the CO(3--2) integrated intensity in the central $\sim 500$ pc. 
The region mainly consists of (i) two spiral arms (labeled 1 and 3 in Figure \ref{fig3}b) that would connect to 
the kpc radius starburst ring \citep[e.g.,][]{1995AJ....110..156Q} and converge to 
(ii) the CND \citep[$r \lesssim 100$ pc region;][]{2013PASJ...65..100I} 
that hosts the putative dusty/molecular torus. 
In addition to the two strong arms, 
there may be another two weak arms \citep[labeled 2 and 4 in Figure \ref{fig3}b, see also][]{2010ApJ...723..767V}. 
Although most of these structures have been identified in 
optical to NIR observations \citep{2001ApJ...553L..35L,2005AJ....130.1472P,2006ApJ...641L..25F,2009ApJ...702..114D,2010ApJ...723..767V,2013ApJ...770L..27F}, 
we map them at cold gas emission 
that would reflect the bulk of the gas mass here. 

Figure \ref{fig3}c shows a close-up view of the CND probed by CO(3--2) emission. 
The spiral arms jointly form 
an arc-like structure with two (north and south) bright knots. 
This arc may correspond to the nuclear star forming region 
inferred by \citet{2005ApJ...624L..13S}. 
We also found a $\sim 10$ pc {\it spatial offset} of the AGN position from the arc. 
Interestingly, a distinctive and compact molecular torus is missing 
at the AGN position of NGC 1097 in this CO(3--2) map. 
This is not due to too high gas excitation conditions at the center 
as we find a similar distribution in an HCN(4--3) 
integrated intensity map as well, which traces 
a factor of $\sim 1000$ denser gas (Kohno et al. in prep.).

\section{Discussion on the torus and summary}\label{sec4}
To discuss torus properties in NGC 1097, 
we first estimate H$_2$ mass ($M_{\rm H_2}$) and the corresponding 
H$_2$ column density ($N_{\rm H_2}$) averaged 
within a single beam (0$\arcsec$.17 $\times$ 0$\arcsec$.14, P.A. = 88$\arcdeg$.8) 
placed at the AGN position. 
The CO(3--2) integrated intensity at this position, 
2.04 $\pm$ 0.21 Jy beam$^{-1}$ km s$^{-1}$, 
was converted to that of CO(1--0) by using 
a CO(3--2)/CO(1--0) brightness temperature ratio 
of 2.33 $\pm$ 0.44 obtained at the central 
7$\arcsec$.9 $\times$ 3$\arcsec$.2 aperture of NGC 1097 \citep{2012ApJ...747...90H}. 
By assuming a CO-to-H$_2$ conversion factor ($X_{\rm CO}$) 
of 0.5 $\times$ 10$^{20}$ cm$^{-2}$ (K km s$^{-1}$)$^{-1}$ \citep{2014A&A...567A.125G}, 
we obtain\footnote{A $\sim 0.3$ dex uncertainty of this $X_{\rm CO}$ is not considered here.} 
$M_{\rm H_2}$ = (5.4 $\pm$ 0.4) $\times$ 10$^4$ $M_\Sun$ 
and $N_{\rm H_2}$ = (1.9 $\pm$ 0.4) $\times$ 10$^{22}$ cm$^{-2}$. 
This $N_{\rm H_2}$ is $\sim 100$ times larger than 
that derived from the X-ray spectral analysis \citep{2006ApJ...643..652N}, 
but it can be a natural consequence of the type-1 (face-on) torus geometry. 
As the $M_{\rm H_2}$ derived with this $X_{\rm CO}$ and that from thermal dust emission 
agrees fairy well for the molecular torus of NGC 1068 \citep{2014A&A...567A.125G,2016ApJ...823L..12G}, 
our use of this factor would be a good starting point for comparisons with NGC 1068. 
This $X_{\rm CO}$ is also comparable to that recommended for 
active environments by \citet{2013ARA&A..51..207B}. 
However, it will be necessary to refine our estimate 
once a better $X_{\rm CO}$ at the nuclear scale is achieved for NGC 1097. 

We then suppose that $M_{\rm H_2}$ and $N_{\rm H_2}$ of NGC 1097 
would be a factor of $\gtrsim 2-3$ smaller than those of 
the molecular torus of NGC 1068 \citep{2016ApJ...822L..10I,2016ApJ...823L..12G}, 
which hosts a much more luminous AGN ($L_{\rm bol} = 2 \times 10^{45}$ erg~s$^{-1}$) 
than NGC 1097 ($L_{\rm bol} = 8.6 \times 10^{41}$ erg~s$^{-1}$). 
As such, this implies that there is less gas (and likely, optically thinner tori) in LLAGNs than in luminous Seyferts. 
We would then actually expect less active nuclear star formation and/or inflows in NGC 1097, 
which will lead to the {\it deficiency of a geometrically thick torus} in this LLAGN 
as compared to luminous AGNs such as NGC 1068 
due to the deficiency of the energy source(s): 
this is an expected trend by the evolutionary torus models (\S 1). 

\begin{figure*}
\epsscale{1.1}
\plotone{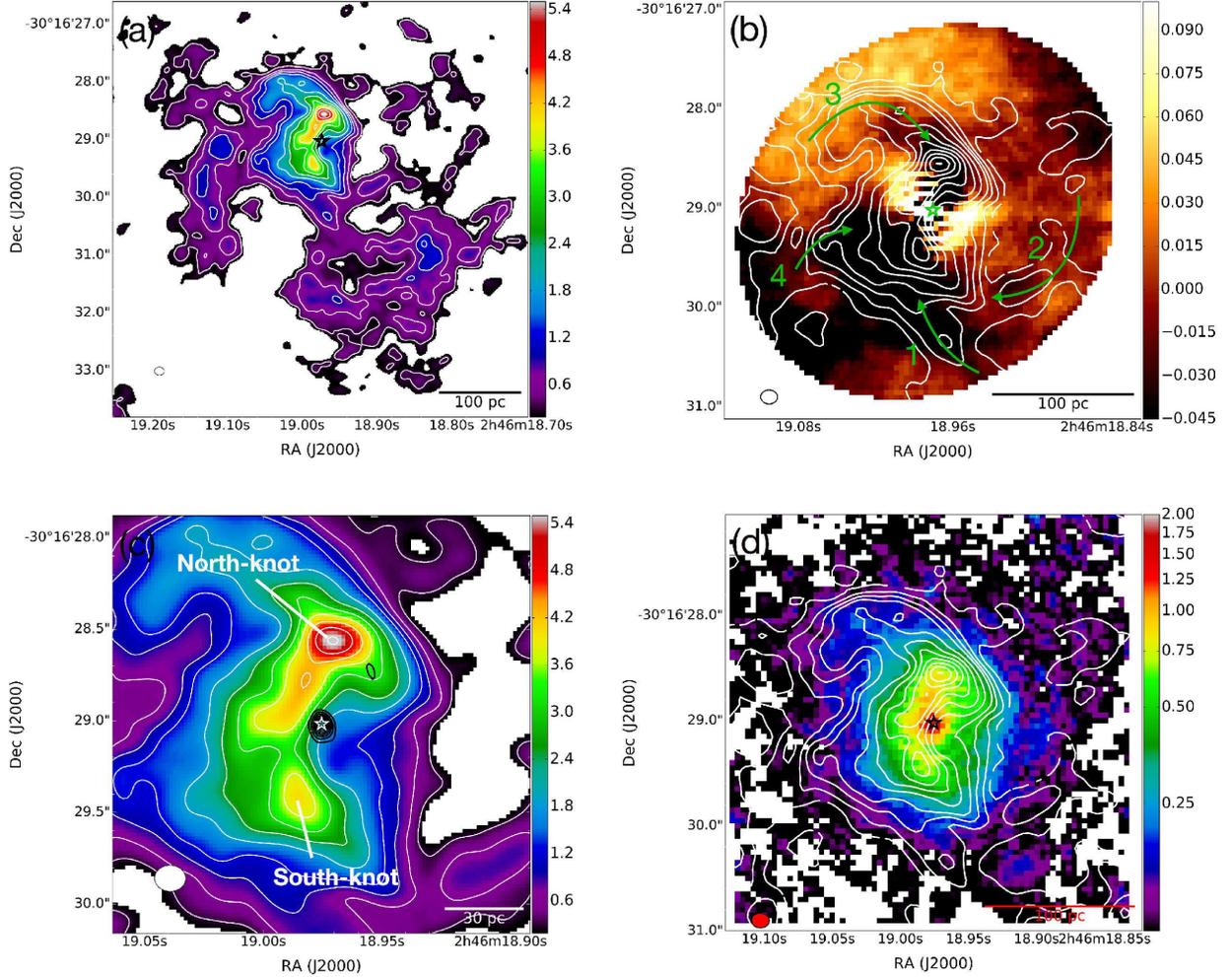}
\caption{
(a) Global spatial distribution of the CO(3--2) integrated 
intensity in the central $\sim 500$ pc of NGC 1097, 
which mainly consists of the CND and two prominent spiral arms. 
Inside the CND, an arc-like structure with two (north and south) knots is recognizable. 
The map shows both the color scale (Jy beam$^{-1}$ km s$^{-1}$ unit) 
and contours (5, 10, 15, 20, 30, $...$, and 100$\sigma$, where 1$\sigma$ = 0.062 Jy beam$^{-1}$ km s$^{-1}$). 
These contour levels are repeated in the other panels. 
(b) CO(3--2) integrated intensity map (contours) overlaid 
on the $K_{\rm s}$-band residual (observed distribution $-$ elliptical model) map 
\citep[arbitrary units;][]{2009ApJ...702..114D}. 
The two prominent spiral arms (labeled 1 and 3), 
as well as another two weak candidate arms (labeled 2 and 4) are highlighted. 
(c) A close-up view of (a) at the innermost $\sim 130$ pc. 
Black contours indicate the 350 GHz continuum emission (Figure \ref{fig1}). 
(d) CO(3--2) integrated intensity map (contours) overlaid 
on the 2.12 $\micron$ H$_2$ integrated intensity map 
\citep[in 10$^{-17}$ W m$^{-2}$ $\micron^{-1}$ units;][]{2009ApJ...702..114D} 
at the innermost $\sim 280$ pc. 
In all panels, the bottom-left ellipse indicates the synthesized beam 
of the CO(3--2) map (0$\arcsec$.17 $\times$ 0$\arcsec$.14, P.A. = 88$\arcdeg$.8), 
and the central star marks the AGN location. 
}
\label{fig3}
\end{figure*}

Another remarkable thing is that the envelope of 
the CND traced by the CO(3--2) emission 
and that traced by the NIR H$_2$ emission 
\citep{2009ApJ...702..114D,2009ApJ...696..448H} 
are almost identical in NGC 1097 at our available resolutions (Figure \ref{fig3}d). 
This indicates that the outer radial extent of 
the molecular CND is unrelated to gas temperature, 
although the vertical height of the CND could differ (see below), 
which is consistent with hydrodynamic simulations of CND-scale 
gas distribution \citep[e.g.,][]{2009ApJ...702...63W}. 
However, we also found that their internal distributions differ significantly: 
while the two knots stand out in the CO(3--2) map, 
the NIR H$_2$ emission is clearly concentrated toward the nucleus, 
likely reflecting complex excitation conditions including fluorescence around the AGN. 
Thus, we should be careful to simply assume a linear conversion 
from NIR H$_2$ line luminosity to CO-based $M_{\rm H_2}$ \citep{2006A&A...454..481M}, 
when using that luminosity as a tracer of molecular mass.

\begin{figure*}
\epsscale{1.1}
\plotone{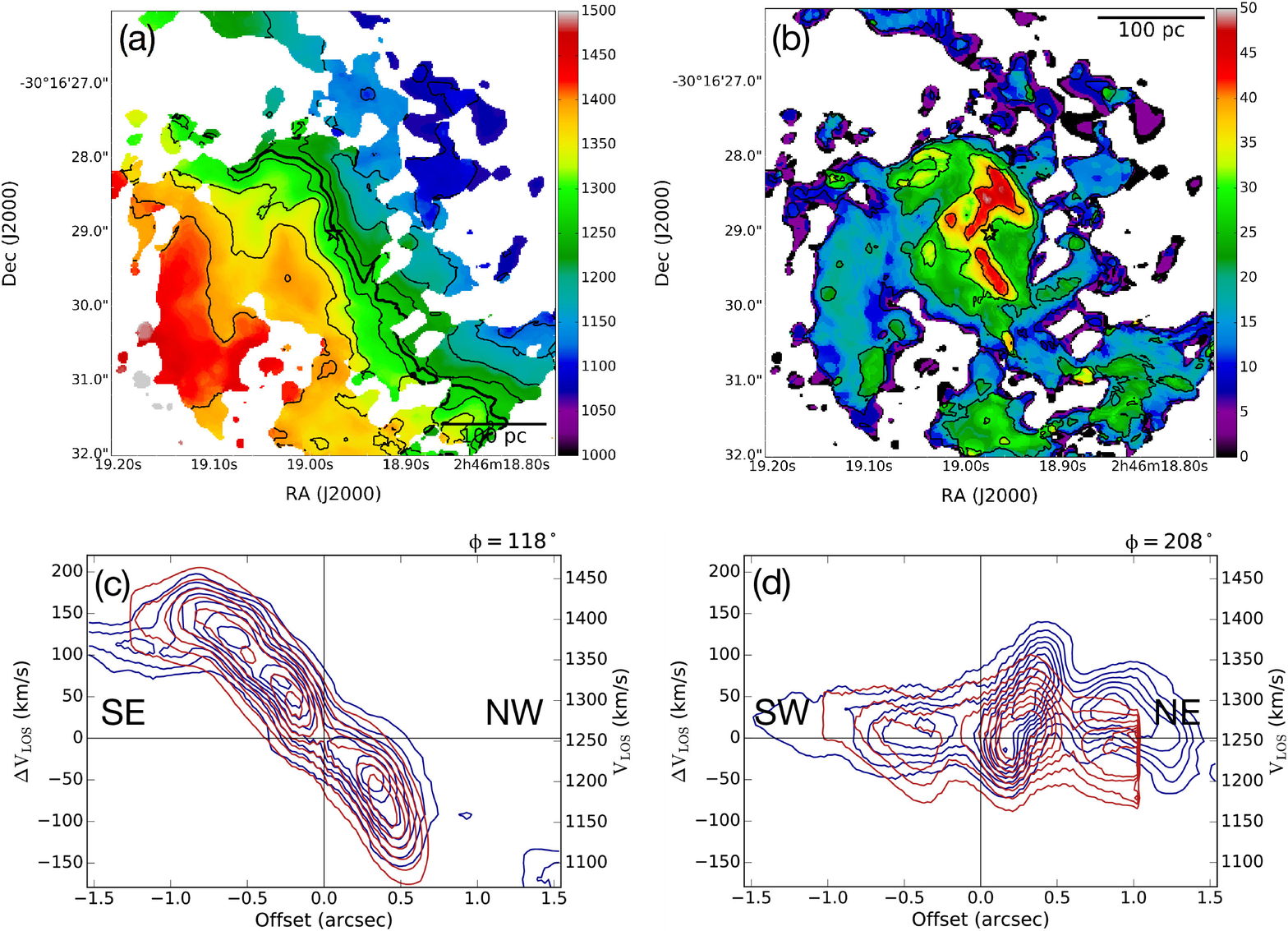}
\caption{
Intensity-weighted (a) mean velocity and (b) velocity dispersion maps 
of the CO(3--2) emission in the central 400 pc of NGC 1097. 
The stars mark the AGN location. The systemic velocity is 1254 km s$^{-1}$ (thick line). 
Thin contours indicate 1080 to 1400 km s$^{-1}$ in steps of 20 km s$^{-1}$ for (a), 
and 10 to 40 km s$^{-1}$ in steps of 10 km s$^{-1}$ for (b), respectively. 
(c)(d) Position-velocity diagrams (PVDs) of the CO(3--2) line 
along the adopted major axis (P.A. = 118$\arcdeg$) and minor axis (P.A. = 208$\arcdeg$) shown in the blue contours. 
The overlaid red contours indicate the PVDs produced by our dynamical model (MODEL-1). 
Both contours are plotted at 5, 10, ..., 40$\sigma$, 
where 1$\sigma$ = 0.84 mJy beam$^{-1}$. 
Due to the inclined geometry of the concentric rings, 
no model component exists at offset $\gtrsim 1\arcsec.1$ along the minor axis. 
}
\label{fig4}
\end{figure*}

Gas dynamics is also the key to study the nature of the torus. 
Figure \ref{fig4}a and b show the observed line-of-sight velocity 
and velocity dispersion maps of the CO(3--2) line in NGC 1097, respectively. 
It is evident that rotation (kinematic P.A. $\simeq 118\arcdeg$) 
dominates the global gas dynamics, 
while streaming motions as well as 
high velocity dispersions along the gas spirals/arc are detected. 

To extract basic beam-deconvolved dynamical information, 
particularly the rotation velocity ($V_{\rm rot}$) and the velocity dispersion ($\sigma$), 
we fitted tilted-rings to these observed maps 
with the $^{\rm 3D}$Barolo code \citep{2015MNRAS.451.3021D}. 
The main parameters are, 
dynamical center, systemic velocity ($V_{\rm sys}$), $V_{\rm rot}$, $\sigma$, 
galactic inclination ($i$), and P.A., all of which can be varied for each ring. 
Our initial fits indicated a dynamical center as the AGN position, 
$V_{\rm sys}$ as 1254 km s$^{-1}$. 
Thus $V_{\rm rot}$, $\sigma$, $i$, and P.A. are free parameters. 
We modeled 19 concentric rings of $\Delta r$ = 0$\arcsec$.05 from the center. 
The $V_{\rm rot}$ and $\sigma$ estimated by the dynamical modeling 
of the NIR H$_2$ emission \citep{2009ApJ...696..448H} were adopted as our initial guesses for the CO(3--2) modeling. 
While this code does not account intrinsically for non-circular motions, 
those can be revealed by subtracting the 2-dimensional rotation map 
from the observed data \citep[e.g.,][]{2016ApJ...823...68S}. 

We first constructed a model with the full resolution CO(3--2) cube 
($0\arcsec.17 \times 0\arcsec.11$, $dV$ = 10 km s$^{-1}$), which we call a MODEL-1. 
Figure \ref{fig4}c and d respectively show the position velocity diagrams (PVDs) 
along the global kinematic major and minor axes, overlaid on the observed data. 
The asymmetric appearances are due to the spatial offset of the AGN from the CO arc. 
The global dynamics can be well reproduced by a combination of gas rotation and dispersion, 
although streaming motions are evident along the minor axis 
(offset = $+0\arcsec.3$ to $+1\arcsec.3$). 

The torus itself (offset = $0\arcsec.0$) does 
not show a significant deviation from this global motion, 
which is a clear contrast to the case of NGC 1068, 
where highly perturbed structures are found in the CO(6--5) velocity field 
\citep{2016ApJ...823L..12G,2016ApJ...829L...7G}. 
This indicates that the molecular torus of NGC 1097 is more quiescent than that of NGC 1068, 
which is further clarified in the following. 
Note that no Keplerian rotation was identified in Figure \ref{fig4}c, 
although the gravitational sphere of influence 
of the previously inferred 1.2 $\times$ 10$^8$ $M_\Sun$ 
black hole \citep[$r \sim 12$ pc or 0$\arcsec$.17;][]{2006ApJ...642..711L} 
should be resolved at our resolution. 
The $M_{\rm BH}$ of NGC 1097 would thus be 
much smaller than previously thought. 
This is consistent with the latest estimate 
($M_{\rm BH} \sim 10^6~M_\Sun$; Hatziminaoglou et al. in prep.) 
based on a kinematic analysis to CO(3--2) data 
also obtained at $\sim 10$ pc resolution. 

\begin{figure*}
\epsscale{1.1}
\plotone{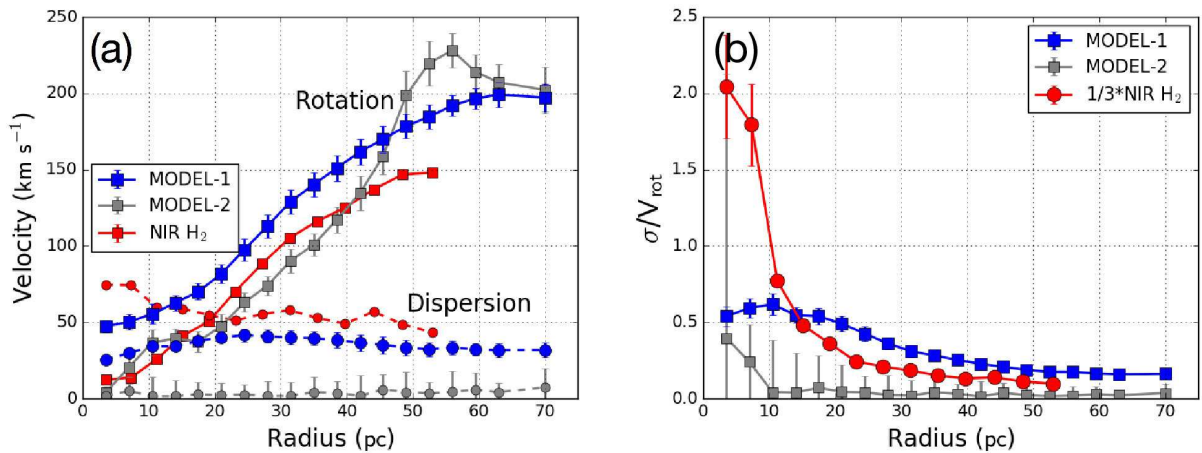}
\caption{
(a) Radial profiles of the rotation velocity ($V_{\rm rot}$; squares) 
and the velocity dispersion ($\sigma$; circles) of the cold molecular gas 
estimated from our tilted-ring modelings to the CO(3--2) velocity field, 
as well as those from the NIR H$_2$ velocity field. 
(b) Radial profiles of the $\sigma/V_{\rm rot}$ ratio 
derived from the CO(3--2) emission line (squares) 
and the NIR H$_2$ emission line (circles; {\it divided by three}). 
In both panels, we plot the data of MODEL-1 and 2 (see \S 4 for details). 
As the MODEL-1 is based on the full resolution CO(3--2) data, we regard it as fiducial. 
}
\label{fig5}
\end{figure*}

Figure \ref{fig5}a shows the resultant radial profiles of 
the decomposed $V_{\rm rot}$ and $\sigma$ of the MODEL-1. 
In terms of the {\it molecular} gas kinematics, 
the absolute values of $V_{\rm rot}$ and $\sigma$ agree relatively well ($\lesssim30\%$) 
with those measured with the NIR H$_2$ line \citep{2009ApJ...696..448H} 
at $r \gtrsim 20$ pc (0$\arcsec$.3). 
Although the resultant $\sigma/V_{\rm rot}$ of the CO(3--2) line 
($\sim 0.2-0.4$; Figure \ref{fig5}b) is a factor of $\sim 1.5$ 
smaller than that of the NIR H$_2$ \citep{2009ApJ...696..448H} there, 
this difference would not be so significant as compared to 
systematic uncertainties of the models according to our experiences. 

On the other hand, the CO(3--2)-based $V_{\rm rot}$ (MODEL-1) 
departs from that derived from the NIR H$_2$, 
particularly at the innermost 15 pc 
(the difference is a factor of $\sim 2-3$). 
We suppose this would be due to 
the mismatched spectral resolutions between 
our CO(3--2) cube (10 km s$^{-1}$) 
and the NIR H$_2$ cube \citep[$\sim 70$ km s$^{-1}$;][]{2009ApJ...696..448H}. 
Indeed, we could achieve a much better agreement ($\lesssim30$\%) in $V_{\rm rot}$ at $r \lesssim50$ pc 
by matching the spectral resolutions (MODEL-2 in Figure \ref{fig5}), while the CO(3--2)-based $\sigma$ 
now becomes one order of magnitude smaller than that derived from the NIR H$_2$. 
Now that the velocity width of the CO(3--2) emission line ($\sim 100$ km s$^{-1}$; Figure \ref{fig4}c) 
is comparable to the matched spectral resolution ($\sim 70$ km s$^{-1}$), 
the CO(3--2) velocity field can be modelled 
well by a combination of $V_{\rm rot}$, $i$, and P.A., without large $\sigma$. 
This in turn suggests that the large $\sigma$ returned from the NIR H$_2$ modeling 
already pointed out a genuine difference in $\sigma$ 
between the CO(3--2) and the NIR H$_2$ lines. 

In either MODEL, the CO(3--2)-based $\sigma/V_{\rm rot}$ 
ratios are much smaller than those derived from the NIR H$_2$ (Figure \ref{fig5}b) at $r \lesssim 20$ pc. 
Thus, while the {\it absolute difference} of the gas dynamics 
probed by each tracer is not well constrained at the center of NGC 1097, 
we would naively imply that cold 
\citep[$\sim 100-300$ K in NGC 1097;][]{2013PASJ...65..100I} 
molecular gas is distributed in a thinner layer 
in and around the torus than the hot ($\sim 1000$ K) component
\footnote{A $\sigma/V_{\rm rot}$ is often used as a proxy of the aspect ratio 
of a disk under the vertical hydrostatic equilibrium.}. 
Regarding the cold molecular torus, 
an estimated scale height with the value of $\sigma/V_{\rm rot} \sim 0.65$ (MODEL-1) 
is then $\sim 4.5$ pc at $r = 7$ pc: 
we regard this MODEL-1 as fiducial 
because it is based on the full resolution CO(3--2) data. 

It is particularly worth noting that the $\sigma/V_{\rm rot}$ ratio of NGC 1097 
at the central $r < 10$ pc ($\lesssim 0.65$; MODEL-1) is significantly smaller than that of 
NGC 1068 \citep[$> 1$;][]{2016ApJ...823L..12G}. 
This is not expected if the disk gravity solely controls the torus thickness, 
considering the higher $M_{\rm dyn}$ in NGC 1068 than in NGC 1097 at their very centers (\S 1). 
Again, the implied less-active star formation and/or inflows from the smaller $M_{\rm H_2}$, 
as well as the orders of magnitude less-luminous AGN activity in NGC 1097 than those in NGC 1068
\footnote{This AGN may host molecular outflow at the torus region, 
which would also thicken the scale height \citep{2016ApJ...829L...7G,2016ApJ...822L..10I}.}, 
would contribute to squash the thick torus in this LLAGN 
\citep[e.g.,][]{2008A&A...491..441V,2009ApJ...702...63W,2012ApJ...758...66W}. 
Note that, very recently, \citet{2017arXiv170605762E} also reported 
a cold molecular gas deficiency at the heart of Cen A. 
As the $L_{\rm bol}$ of Cen A is $\sim 3 \times 10^{42}$ erg s$^{-1}$ \citep[LLAGN;][]{2010MNRAS.402..724P}, 
their findings would also fit the evolutionary torus models. 

In summary, although the actual physical process remains unclear, 
we support the theoretical prediction 
that geometrically and optically thick tori 
will gradually become deficient as AGNs 
evolve from luminous Seyferts to more quiescent LLAGNs. 
It is mandatory to enlarge the sample with 
$\lesssim 10$ pc resolution cold gas measurements to confirm this trend.

\acknowledgments
We thank the anonymous referee for his/her careful reading and 
useful comments which greatly improved this paper. 
This paper makes use of the following ALMA data: 
ADS/JAO.ALMA\#2015.1.00126.S and \#2012.1.00187.S. 
ALMA is a partnership of ESO (representing its member states), 
NSF (USA) and NINS (Japan), together with NRC (Canada), 
NSC and ASIAA (Taiwan), and KASI (Republic of Korea), 
in cooperation with the Republic of Chile. 
The Joint ALMA Observatory is operated by ESO, AUI/NRAO and NAOJ. 
We sincerely thank E. K. S. Hicks for kindly providing the NIR H$_2$ kinematic data. 
T.I., N.K., K.W., and K.K. are supported by JSPS KAKENHI Grant Number 
17K14247, 16K17670, 16H03959, and 25247019, respectively. 
T.I. was supported by the ALMA Japan Research Grant 
of NAOJ Chile Observatory, NAOJ-ALMA-170.

\end{document}